\documentclass[12pt]{article}
\usepackage{amsmath,amssymb}
\usepackage[dvips]{graphicx}

\newcommand{\dd}{\ddots}
\newcommand{\pp}{\!\!+\!\!}
\newcommand{\mm}{\!\!-\!\!}
\newcommand{\p}{\!+\!}
\newcommand{\m}{\!-\!}

\begin{document}



\begin{titlepage}

  \begin{normalsize}
  \begin{flushright}
    YITP-01-20\\
    hep-th/0103146\\
    March 2001
  \end{flushright}
  \end{normalsize}
  
  \vspace{1cm}
  
  \begin{Large}
  \begin{center}
  \bf
    Torus-like Dielectric D2-brane
  \end{center}
  \end{Large}
  
  \vspace{1cm}

\begin{center}
   Yoshifumi Hyakutake
   \footnote{E-mail :hyaku@yukawa.kyoto-u.ac.jp}
   \\
   \vspace{6mm}
   {\it Yukawa Institute for Theoretical Physics, Kyoto University} 
   \\
   {\it Sakyo-ku, Kyoto 606-8502, Japan}
\end{center}

\vspace{2cm}
\begin{center}
  \large{ABSTRACT}
\end{center}

\begin{quote}
\begin{normalsize}

\end{normalsize}
\end{quote}
We find new solutions corresponding to torus-like generalization
of dielectric D2-brane from the viewpoint of D2-brane action and 
$N$ D0-branes one. These are meta-stable and would decay to the spherical 
dielectric D2-brane.
\end{titlepage}

\setlength{\baselineskip}{0.64cm}

\section{Introduction and Preliminaries}

Type II superstring theories possess stringy solitons called Dirichlet-branes
(D-branes) and they have been attracting our attention as a clue to resolve 
the non-perturbative aspects. 
As originally defined, a D$p$-brane is a $(p\!+\!1)$-dimensional hypersurface
in 10-dimensional space-time on which open strings are allowed to end and
the massless modes of the open strings form a multiplet of supersymmetric 
$U(1)$ gauge theory, a vector field $A_a$, $(9\!-\!p)$ real scalar 
fields $\Phi^i$ and their superpartners\cite{Pol}. 
We will use $a,b$ for world-volume 
indices of D$p$-branes, $i,j,k$ for transverse indices of that and $\mu,\nu$ 
for space-time indices. 

The bosonic part of the low energy theory on a single D$p$-brane world-volume
is described by an abelian Born-Infeld action of the form\cite{Lei}
\begin{alignat}{3}
  \mathcal{S}_{\text{BI}} &= - T_p \!\int\!\! 
  d\!\stackrel{p+1}{\!\!\!\sigma} \! e^{-\phi}
  \sqrt{-\det (P[G_{ab}+B_{ab}] + \lambda F_{ab})} \;, \label{eq:BI}
\end{alignat}
where $T_p = 2\pi/(2\pi\ell_s)^{p+1} g_s$ is the D$p$-brane tension and
$\lambda = 2\pi\ell_s^2$.
$F_{ab}$ is a field strength of the gauge field on the world-volume theory.
A dilaton $\phi$, a metric $G_{\mu\nu}$ and an antisymmetric tensor 
$B_{\mu\nu}$, which are generically called Neveu-Schwartz fields, are fields 
in the space-time, so they appear in the above action as pullbacks to the 
world-volume.
We'll use the symbol $P[\cdots]$ to clarify the pullbacks, for example, 
$P[G_{ab}] = \frac{\partial x^\mu}{\partial \sigma^a}
\frac{\partial x^\nu}{\partial \sigma^b} G_{\mu\nu}$.
To simplify the action we often employ the static gauge
\begin{alignat}{3}
  \begin{cases}
    x^a = \sigma^a ,\quad & a=0,1,\cdots,p \;, \\
    x^i = \lambda \Phi^i(x^a) ,\quad & i=p\!+\!1,\cdots,9 \;.
  \end{cases}
\end{alignat}
In this case $P[G_{ab}]$ is written as
\begin{alignat}{3}
  P[G_{ab}] &= G_{ab} 
  + \lambda (\partial_a \Phi^i G_{ib} + \partial_b \Phi^i G_{ai})
  + \lambda^2 \partial_a \Phi^i \partial_b \Phi^j G_{ij}.
\end{alignat}

One of the important features of the abelian Born-Infeld action 
is that it can capture
some geometrical informations about the D$p$-brane and strings attached to or
dissolved in it\cite{CM,Gib,SAL,Emp,PTMM}. 
For instance, in the case of a D3-brane world-volume theory,
a point charge Coulomb gauge field become a BPS
saturated solution if one of the scalar fields, 
say $\Phi^9$, is exited. The excitation of $\Phi^9$ implies that the shape of 
the D3-brane is like a spike and, indeed, this configuration is identified 
with a fundamental string attached to the D3-brane\cite{CM}.

In addition to the massless NS spectrums, closed strings in type II 
superstring theories contain Ramond-Ramond massless spectra. A D-brane 
carries an R-R charge and couples to R-R potentials as\cite{ML,Dou,GHM}
\begin{alignat}{3}
  \mathcal{S}_{\text{CS}} = \mu_p \!\int_{p+1} P \big[\sum_{n \leq p\!+\!1} 
  C^{(n)}e^B \big] e^{\lambda F} . \label{eq:Che}
\end{alignat}
This is an abelian Chern-Simons action for the single D$p$-brane.
Here $\mu_p = T_p$ is the R-R charge of the D$p$-brane and 
$C^{(n)}$ denotes the R-R $n$-form potential. $n$ takes odd $(1,3,5,7,9)$ for 
type IIA and even $(0,2,4,6,8,10)$ for type IIB.
In the presence of non-zero R-R field strength $G^{(n+1)}=dC^{(n)}$, the 
abelian Chern-Siomons action gives some contribution to potential energy
of the D$p$-brane world-volume theory, 
which implies existences of new solutions. For example, 
a spherical D2-brane, in the $x^1,x^2,x^3$ directions, 
with $N$ magnetic fluxes is usually unstable against decaying to $N$ D0-branes.
In the background of constant $G^{(4)}_{0123} = -4h$, however, 
this spherical D2-brane expanding in the $x^1,x^2,x^3$ directions with a
radius $h\lambda N$ becomes globally stable.
This is so-called a dielectric effect or spherically dielectric 
D2-brane\cite{Mye}.

Low energy world-volume action for the single D$p$-brane is realized by
the abelian Born-Infeld action $(\ref{eq:BI})$ plus 
the abelian Chern-Simons action $(\ref{eq:Che})$. 
Now we move on to a world-volume action for $N$ coincident
D$p$-branes. If $N$ parallel D$p$-branes close to each other, the ground
state modes of open strings stretched between these D$p$-branes 
become massless and
therefore world-volume gauge symmetry is enhanced to $U(N)$\cite{Wit}.
To obtain the world-volume action of the $N$ coincident D$p$-branes, 
in the actions $(\ref{eq:BI})$ and $(\ref{eq:Che})$,
we would modify $A_a$ and $\Phi^i$ as adjoint representations of $U(N)$
Lie algebra and take a trace of the integrand. 
The action obtained like this way, however, 
is not consistent with T-duality. The correct one is given by Myers et al.
as follows\cite{Mye,CMT,Tse,TR}.

The non-abelian Born-Infeld action for the $N$ coincident D$p$-branes,
which is consistent with T-duality, is modified into the form
\begin{alignat}{3}
  \mathcal{S}_{\text{BI}} &= - T_p && \!\!\int\!\! 
  d\!\stackrel{p+1}{\!\!\!\sigma} \! \text{STr}
  \bigg( \!\! e^{-\phi} \!\! \sqrt{\!-\det \big( P[E_{ab}] \!+\! 
  P[E_{ai} ({(Q^{-1})^i}_j \!-\! {\delta^i}_j)E^{jk}E_{kb}] 
  \!+\! \lambda F_{ab} \big) \det \big({Q^i}_j \big)} \bigg), \label{eq:nBI}
\end{alignat}
where $E_{\mu\nu} \!=\! G_{\mu\nu} \!+\! B_{\mu\nu}$ and 
${Q^i}_j \!=\! {\delta^i}_j \!+\! i\lambda [\Phi^i,\Phi^k]E_{kj}$.
The partial derivatives $\partial_a \Phi^i$ are changed to covariant 
derivatives $D_a \Phi^i \!\equiv\! \partial_a \Phi^i \!+\! i[A_a,\Phi^i]$.
STr represents a symmetrized trace, that is, $U(N)$ Lie algebra indices of 
$F_{ab}$, $D_a \Phi^i$ and $i[\Phi^i,\Phi^j]$ are symmetrized in the trace.
By setting $\phi=B_{\mu\nu}=0$ and expanding around 
$G_{\mu\nu} = \eta_{\mu\nu}$, the non-abelian Born-Infeld action 
is approximated of the form
\begin{alignat}{3}
  \mathcal{S}_{\text{BI}} &\sim - N T_p V_{p+1} - \frac{\lambda^2}{4} T_p
  \!\!\int\!\! d\!\stackrel{p+1}{\!\!\!x} \! \text{Tr}
  \Big( F^{ab}F_{ab} + 2(D^a \Phi^i)(D_a \Phi^i) - [\Phi^i,\Phi^j]
  [\Phi^i,\Phi^j] \Big). \label{eq:Lap}
\end{alignat}
This action is also obtained by the dimensional reduction of the bosonic
part of 10-dimensional $U(N)$ SYM theory to $(p+1)$-dimensional theory.
It is easy to see that the potential term $-[\Phi^i,\Phi^j]^2$ is
characteristic of
the non-abelian case, which originally comes from the term $\sqrt{\det Q}$.
This potential term reaches to global minima 
when all $\Phi^i$ are simultaneously
diagonalized. The diagonal elements of each $\Phi^i$ multiplied by
$\lambda$ represent the positions of the individual D$p$-branes 
in the $x^i$ direction.

The non-abelian Chern-Simons action for the $N$ coincident D$p$-branes
, which is consistent with T-duality, is modified into the form
\begin{alignat}{3}
  \mathcal{S}_{\text{CS}} &= \mu_p \!\int_{p+1} \text{STr} \Big(
  P \big[e^{i \lambda {\rm i}_\Phi {\rm i}_\Phi} 
  \big(\sum_n C^{(n)}e^B \big) \big] e^{\lambda F} \Big), \label{eq:nCh}
\end{alignat}
where $n$ takes odd for type IIA and even for type IIB. Here $\rm{i}_\Phi$ 
denotes the interior product defined as
\begin{alignat}{3}
  ({\rm i}_\Phi {\rm i}_\Phi)^l C^{(n+2l)} = \frac{1}{2^l n!} \Big( 
  \prod_{s=1}^l [\Phi^{j_{2s}},\Phi^{j_{2s-1}}] \Big)
  C^{(n+2l)}_{j_1 j_2 \cdots j_{2l-1} j_{2l}
  \mu_1 \cdots \mu_n} dx^{\mu_1} \wedge \cdots \wedge dx^{\mu_n} .
\end{alignat}
The remarkable property of the non-abelian Chern-Simons action is that 
there are coupling terms to the R-R potential $C^{(n)}$ with $n \geq p+3$. 
Because of this fact, a non-zero constant R-R field strength $G^{(p+4)}$ 
generates a new potential term for the $N$ coincident D$p$-branes 
world-volume theory.
For instance, in the presence of the constant $G^{(4)}_{0123} = -4h$, 
the configurations of $N$ static D0-branes, i.e. all simultaneously 
diagonalized $\Phi^i$, are no longer globally stable. 
The potential energy reaches to global minima when $\frac{\Phi^1}{2h}$, 
$\frac{\Phi^2}{2h}$ and $\frac{\Phi^3}{2h}$ satisfy the commutation
relation of the $SU(2)$ Lie algebra.
These solutions are known as fuzzy two-spheres. If $\frac{\Phi^1}{2h}$, 
$\frac{\Phi^2}{2h}$ and $\frac{\Phi^3}{2h}$ belong to the $N$ dimensional
irreducible representations of $SU(2)$ Lie algebra, the radius of the
fuzzy two-sphere is almost equal to $h\lambda N$.
This is the dielectric effect viewed from the $N$ D0-branes world-volume 
theory, often called Myers effect.

In this paper we explore torus-like configurations of the dielectric D2-brane
from the two alternative viewpoint of 
the single D2-brane action or the $N$ D0-branes one\footnote{ In 
refs.\cite{BJL} and \cite{BJL2},
a dielectric D5-brane with the topology $\mathbb{R}^4 \times T^2$
is considered from a viewpoint of certain deformations of
$\mathcal{N}=4$ super Yang-Mills theory in four dimensions.}.
Motivations are as follows. Firstly, consider a situation where the
spherical D2-brane (the fuzzy two-sphere) is expanding 
in the $x^1,x^2,x^3$ directions and the
$x^3$ direction is compactified on a circle with circumference $2\ell$. 
The radius of the spherical D2-brane (the fuzzy two-sphere) is given 
by $h\lambda N$ in the background of constant $G^{(4)}_{0123}=-4h$ flux.
Therefore, if $h\lambda N$ is bigger than $\ell$, the spherical D2-brane 
(the fuzzy two-sphere) would be
deformed to a torus-like configuration via tachyon condensation process.
Notice that one of two cycles of this torus wraps the $x^3$ direction
but another does not wind any circle of the target space.
It might difficult to pursuit the tachyon condensation process directly. 
However, knowledges of possible torus-like configurations of the dielectric 
D2-brane (the $N$ D0-branes) would tell us what happens after 
the tachyon condensation. 
The another reason is that the torus-like dielectric D2-brane
($N$ D0-branes) would be possible to regard 
as a non-BPS D-string which wraps the $x^3$ direction.
A non-BPS D-string can be constructed from the D2-$\overline{\text{D}2}$
system by the condensation of a real tachyonic field.
To make it more clear, consider the case where D2 and $\overline{\text{D}2}$
are extending in the $x^2,x^3$ directions and separated by the string scale
in the $x^1$ direction. If the tachyon condensation starts
from $x^2=\pm\infty$ and proceed step by step toward $x^2=0$, 
a torus-like D2-brane can be obtained. A non-BPS D-string is identified
to this torus-like D2-brane if the cycle, which doesn't wind any circle
of the target space, shrinks almost to zero. 
The constant R-R flux, however, might
puff up this shrinking cycle. It is interesting if we can construct
a non-BPS D-string via $N$ D0-branes.

In section \ref{sec:Rev} we review the dielectric effect briefly.
In section \ref{sec:Tor} the torus-like configurations of the 
dielectric D2-brane are 
obtained as classical solutions of equation of motion for the D2-brane
world-volume action.
We try to reproduce the same configurations from the viewpoint of the
$N$ D0-branes action in section \ref{sec:TorMye}.
In section \ref{sec:Tac} we make some discussions.

\vspace{1cm}
\section{Brief Review of the Dielectric Effect}
\label{sec:Rev}

The world-volume theory of $N$ coincident D$p$-branes is described 
by non-abelian Born-Infeld action plus non-abelian Chern-Simons action, 
which are consistent with T-duality\cite{Mye,CMT,Tse,TR}.
The remarkable fact is that, in the presence of constant R-R 4-form
field strength, the $N$ coincident D0-branes are unstable against expanding
into a spherical D2-brane\cite{Mye}. In this section we briefly review
this dielectric effect.

\subsection{The Dielectric Effect from the D2-brane Action}
\label{subsec:sph}

In this subsection we shortly see the dielectric effect
from the viewpoint of a single D2-brane world-volume theory.
The system considered here is the spherical 
D2-brane with $N$ magnetic fluxes in the presence of constant 
R-R flux $G^{(4)}_{0123}=-4h$. The space-time flat metric is taken to be
\begin{alignat}{3}
  ds^2 = -dt^2 + dr^2 + r^2 d\theta^2 + r^2 \sin^2 \theta \, d\phi^2
  + \sum_{s=4}^{9} (dx^s)^2.
\end{alignat}
Here the $x^1$, $x^2$ and $x^3$ directions are expressed by the spherical
coordinates as 
\begin{alignat}{3}
  x^1 = r \sin \theta \cos \phi ,\quad x^2 = r \sin \theta \sin \phi
  ,\quad x^3 = r \cos \theta .
\end{alignat}
Instead of the usual static gauge, it is useful to take
the world-volume coordinates of the spherical D2-brane as
$\sigma^0 \!=\! t$, $\sigma^1 \!=\! \theta$, $\sigma^2 \!=\! \phi$. 
Suppose that the
position of the spherical D2-brane is at $r \!=\! R$ and $x^s \!=\! 0$.
The R-R flux on the spherical coordinates is $G^{(4)}_{tr\theta\phi}=
- 4h r^2 \sin\theta$, or $C^{(3)}_{t\theta\phi} = \frac{4}{3}h r^3 \sin\theta$ 
in terms of the R-R potential. The field strength of the gauge field
is given by $F_{\theta\phi} = \frac{1}{2}N \sin\theta$ and other fields are
set to be zero.

In these backgrounds, the world-volume action of the spherical D2-brane
is given by
\begin{alignat}{3}
  {\mathcal S}_{\text{D2}} &= - T_2 \int \!\! dtd\theta d\phi
  \sqrt{-\det (P[G_{ab}] + \lambda F_{ab})} + \mu_2 \int P[C^{(3)}] \notag
  \\
  &= -4\pi T_2 \int \!\! dt \bigg( \sqrt{R^4 + \frac{\lambda^2 N^2}{4}}
  - \frac{4h}{3} R^3 \bigg).
\end{alignat}
Thus, the potential energy is expressed by a function of $R$ as
\begin{alignat}{3}
  V_{\text{D2}}(R) = \sqrt{\big(4\pi R^2 T_2 \big)^2 + \big(N T_0 \big)^2}
  - \frac{16\pi}{3}h T_2 R^3 . \label{eq:pot}
\end{alignat}
The first term is just the square root of the sum of 
the spherical D2-brane mass squared
and $N$ D0-branes mass squared. If we assume $4\pi R^2 T_2 \ll N T_0$, this
potential energy can be approximated as
\begin{alignat}{3}
  V_{\text{D2}}(R) &\sim N T_0 + \frac{2 T_0}{\lambda^2 N} R^4
  - \frac{8 h T_0}{3 \lambda} R^3 . \label{eq:app}
\end{alignat}
This approximated potential energy takes extrema at $R\!=\!0, h \lambda N$. 
The potential energies at these points are
$V_{\text{D2}}(0) \!=\! N T_0$ and $V_{\text{D2}}(h\lambda N)\!=\!
N T_0 - \frac{2}{3} NT_0 (h^2 \lambda N)^2$ respectively, so the spherical 
D2-brane with $R\!=\!h\lambda N$ is stable. 
In this case the above assumption insists $h^2 \lambda N \ll 1$.

Before proceeding to the next subsection we should estimate a region
where our calculations so far are reliable. If we take a typical length
of the system to be $R_{\text{sys}}$, 
dimensionless square of gauge coupling constant on the 
D2-brane is given by $g_{\text{gauge}}^2 R_{\text{sys}} \sim 
g_s R_{\text{sys}}/\ell_s$ and that of gravitational coupling constant becomes
$\kappa^2/R_{\text{sys}}^8 \sim \ell_s^8 g_s^2/R_{\text{sys}}^8$. 
The Born-Infeld action is valid where
$g_{\text{gauge}}^2 R_{\text{sys}}$ is fixed and the gravitational 
interactions are neglected,
i.e., in the decoupling limit $\ell_s \ll R_{\text{sys}}$ and $g_s \ll 1$.
So in the case of $R \!=\! h\lambda N$, we should require 
$\ell_s \ll h\lambda N$.

\subsection{Myers Effect}
\label{subsec:Mye}

In this subsection we study
the dielectric effect from the viewpoint of $N$ coincident D0-branes.
The world-volume action employed here is
(\ref{eq:Lap}) plus (\ref{eq:nCh}). 
The $\Phi^4,\cdots,\Phi^9$ are set to be zero.
Now the world-volume action of the $N$ coincident D0-branes in the background
of constant R-R flux $G^{(4)}_{0123} = -4h$ is given by
\begin{alignat}{3}
  {\mathcal S}_{\text{$N$D0}} &\sim
  -\! N T_0 V_1 \!+\! \frac{\lambda^2}{4} T_0 \!\!\int\!\! dt \, \text{Tr}
  \Big( 2(\partial_t \Phi^i)^2 \!+\! [\Phi^i,\Phi^j]
  [\Phi^i,\Phi^j] \Big) \!-\! \frac{2i h\lambda^2}{3} \mu_0 \, \epsilon_{ijk} 
  \!\!\int\!\! dt \, \text{Tr} \Big( \Phi^i[\Phi^j,\Phi^k] \Big),
\end{alignat}
where $i,j,k$ run $1,2,3$.
This action is valid when the commutators $i\lambda[\Phi^i,\Phi^j]$ 
are small enough. The potential energy can be written as
\begin{alignat}{3}
  V_{\text{$N$D0}} = NT_0 - \frac{\lambda^2}{4}T_0 \text{Tr}
  \Big( [\Phi^i,\Phi^j][\Phi^i,\Phi^j] \Big) + \frac{2ih \lambda^2}{3}\mu_0
  \, \epsilon_{ijk} \text{Tr} \Big( \Phi^i[\Phi^j,\Phi^k] \Big) .
  \label{eq:pot2}
\end{alignat}
And in the static case, the equation of motion is obtained by varing
the potential energy,
\begin{alignat}{3}
  \frac{\delta V_{\text{$N$D0}}}{\delta \Phi^i} = 
  \lambda^2 T_0 \, [[\Phi^i,\Phi^j],\Phi^j] + 2ih \lambda^2 \mu_0 \,
  \epsilon_{ijk} [\Phi^j,\Phi^k] = 0 . \label{eq:eom}
\end{alignat}
All simultaneously diagonalized $\Phi^i$ are one of the solutions of this
equation of motion, where the potential energy is equal to $NT_0$.
On the other hand, we can find more nontrivial solutions which satisfy
the relation of $SU(2)$ Lie algebra
\begin{alignat}{3}
  \Big[ \frac{\Phi^i}{2h},\frac{\Phi^j}{2h} \Big] 
  &= i \epsilon_{ijk} \frac{\Phi^k}{2h} .
\end{alignat}
The non-commutative geometry which satisfies the above relation is often 
called fuzzy two-sphere. If all $\Phi^i/2h$ belong to $N$ 
dimensional irreducible representations of $SU(2)$ Lie algebra, 
the radius of the fuzzy two-sphere is given by
\begin{alignat}{3}
  R_{\text{fuzzy}} \equiv \sqrt{\frac{1}{N} \text{Tr} \big((\lambda \Phi^i)^2 
  \big)} = h \lambda N \sqrt{1-\frac{1}{N^2}} \;,
\end{alignat}
and the potential energy can be estimated as
\begin{alignat}{3}
  V_{\text{$N$D0}} &= NT_0 - \frac{2}{3} NT_0 (h^2 \lambda N)^2
  \Big( 1 - \frac{1}{N^2} \Big) .
\end{alignat}
This potential energy is less than that of the all simultaneously diagonalized
solutions and this means that the fuzzy two-sphere is stable. 
If the $N$ is large, the radius and the potential energy of the fuzzy 
two-sphere match up to those of the spherical D2-brane calculated in the
previous subsection up to $1/N^2$ correction.

As a preparation for a later section, we examine the geometry of the fuzzy 
two-sphere in detail. By using variables
\begin{alignat}{3}
  J_+ = \frac{\Phi^1}{2h} + i \frac{\Phi^2}{2h} \;,\quad
  J_- = \frac{\Phi^1}{2h} - i \frac{\Phi^2}{2h} \;,\quad
  J_3 = \frac{\Phi^3}{2h} \;,
\end{alignat}
the equation of motion (\ref{eq:eom}) is rewritten as
\begin{alignat}{3}
  &\big[[J_+,J_-],J_+ \big] + 2\big[[J_+,J_3],J_3 \big] + 4[J_+,J_3] = 0 
  \;, \label{eq:eom2}
  \\
  &\big[[J_+,J_3],J_- \big] + \big[[J_-,J_3],J_+ \big]  + 2[J_+,J_-] = 0 
  \;. \notag
\end{alignat}
In the case of the fuzzy two-sphere, $J_+$ and $J_-$ are ladder operators 
of $SU(2)$ Lie algebra and $J_3$ is the Cartan subalgebra. 
The explicit matrix form of $J_3$ in the $N$ dimensional irreducible 
representation is given by
\begin{alignat}{3}
  J_3^{\text{irr}} = 
  \begin{pmatrix}
    \; \frac{N-1}{2} & & & & \;\\
    \: & \frac{N-3}{2} & & & \;\\
    \; & & \ddots & & \;\\
    \; & & & -\frac{N-3}{2} & \;\\
    \; & & & & -\frac{N-1}{2} \;
  \end{pmatrix} , \label{eq:J_3}
\end{alignat}
where the superscript is to denote the irreducibility.
Each diagonal element multiplied by $2h\lambda$ would be interpreted as 
the position of each D0-brane in the $x^3$ direction.
Here we call the D0-brane corresponding to the $(m,m)$ element as
the $m$th D0-brane.
The explicit matrix forms of $J_+$ and $J_-$ belonging to the  $N$ dimensional
irreducible representations are given by
\begin{alignat}{3}
  J_+^{\text{irr}} = 
  \begin{pmatrix}
    \; 0 & \sqrt{N\!\!-\!\!1} & & & \;\\
    \: & 0 & \!\!\!\!\!\!\sqrt{2(N\!\!-\!\!2)}\!\!\!\! & & \;\\
    \; & & \ddots & & \;\\
    \; & & & 0 & \sqrt{N\!\!-\!\!1} \;\\
    \; & & & & 0 \;
  \end{pmatrix} \;,\;
  J_-^{\text{irr}} = 
  \begin{pmatrix}
    \; 0 & & & & \;\\
    \: \sqrt{N\!\!-\!\!1} & 0 & & & \;\\
    \; & \!\!\!\!\!\!\sqrt{2(N\!\!-\!\!2)}\!\! & \ddots & & \;\\
    \; & & & 0 & \;\\
    \; & & & \!\!\sqrt{N\!\!-\!\!1} & 0 \;
  \end{pmatrix}. \label{eq:J_+-}
\end{alignat}
The off-diagonal elements $(m,m\p 1)$ or $(m\p 1,m)$ multiplied by $2h\lambda$
would represent the extension, to the radial direction of 
the $x^1$-$x^2$ plane, of the cell between the $m$th D0-brane 
and the $(m\p 1)$th D0-brane.
The factor $2h\lambda$ is defined by the fact that 
$\text{Tr} \big((\lambda \Phi^1)^2\big) + \text{Tr} 
\big((\lambda \Phi^2)^2 \big)$ is equal to 
$(2h\lambda)^2 \text{Tr}\big(J_+ J_- \big)$.

On the arguments so far, we chose the irreducible representations of
$SU(2)$ Lie algebra as the solution of (\ref{eq:eom2}).
But reducible representations of $SU(2)$ can also be the solutions.
For example $2N \!\times\! 2N$ matrices,
\begin{alignat}{3}
  J_3 = 
  \begin{pmatrix}
    J_3^{\text{irr}} \!+\! c \mathbf{1}_{N\times N} & 0 \\
    0 & J_3^{\text{irr}} \!-\! c \mathbf{1}_{N\times N}
  \end{pmatrix}
  ,\quad
  J_+ = 
  \begin{pmatrix}
    J_+^{\text{irr}} & 0 \\
    0 & J_+^{\text{irr}}
  \end{pmatrix}
  ,\quad
  J_- = 
  \begin{pmatrix}
    J_-^{\text{irr}} & 0 \\
    0 & J_-^{\text{irr}}
  \end{pmatrix}
  ,
\end{alignat}
become a solution and represent two fuzzy two-spheres located
at $x^3=2h \lambda c$ and $x^3=-2h \lambda c$ respectively. 
The potential energy of this solution is
higher than that of a fuzzy two-sphere which is constructed by $2N$ dimensional
irreducible representations, but lower than that of $N$ coincident D0-branes.

Since each component of the commutators $i\lambda [\Phi^i,\Phi^j]$
is of the order of $h^2\lambda N$, our calculations are reliable
when a relation $h^2\lambda N \ll 1$ holds.

\vspace{1cm}
\section{A Torus-like Dielectric D2-brane}
\label{sec:Tor}

In this section we find out configurations of a torus-like D2-brane with 
$N$ magnetic fluxes in the presence of constant R-R flux 
$G^{(4)}_{0123} = -4h$. The $x^3$ direction is compactified by a circle
with the circumference $2\ell$. One of two cycles of the torus-like D2-brane 
wraps the $x^3$ direction, but another doesn't wind any circle of the target
space. This will be an extension of the spherical D2-brane solution.

\begin{figure}[tb]
\begin{center}
  \begin{picture}(360,140)
    \thicklines
    \put(30,20){\line(0,1){120}} 
    \put(5,37){$-2\ell$}
    \put(10,57){$-\ell$}
    \put(18,77){0}
    \put(19,97){$\ell$}
    \put(14,117){$2\ell$}
    \put(27,143){$x^3$}
    \put(27,40){\line(1,0){6}}
    \put(27,60){\line(1,0){6}}
    \put(27,80){\line(1,0){6}}
    \put(27,100){\line(1,0){6}}
    \put(27,120){\line(1,0){6}}
    \put(110,40){\circle{30}} 
    \put(110,80){\circle{30}}
    \put(110,120){\circle{30}}
    \put(103,0){(a)}
    \put(205,40){\circle{40}} 
    \put(205,80){\circle{40}}
    \put(205,120){\circle{40}}
    \put(197,0){(b)}
    \qbezier(285,20)(285,25)(280,30) 
    \qbezier(280,30)(273,40)(280,50)
    \qbezier(280,50)(290,60)(280,70)
    \qbezier(280,70)(273,80)(280,90)
    \qbezier(280,90)(290,100)(280,110)
    \qbezier(280,110)(273,120)(280,130)
    \qbezier(280,130)(285,135)(285,140)
    \qbezier(315,20)(315,25)(320,30)
    \qbezier(320,30)(327,40)(320,50)
    \qbezier(320,50)(310,60)(320,70)
    \qbezier(320,70)(327,80)(320,90)
    \qbezier(320,90)(310,100)(320,110)
    \qbezier(320,110)(327,120)(320,130)
    \qbezier(320,130)(315,135)(315,140)
    \put(293,0){(c)}
  \end{picture}
  \caption{Configurations of the dielectric D2-brane.
  The $x^3$ direction is compactified with the circumference $2\ell$.
  (a) $h\lambda N \!<\! \ell$. (b) $h\lambda N \!\sim\! \ell$.
  (c) $h\lambda N \!>\! \ell$.}
  \label{fig:cyl}
\end{center}
\end{figure}
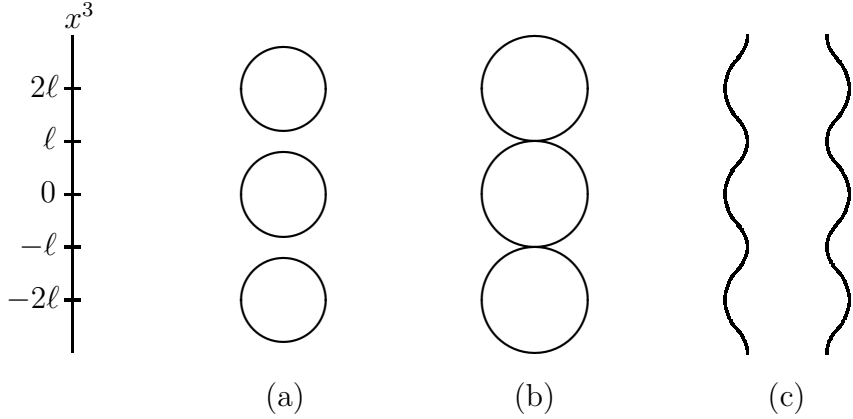

Let us consider a case where the center of the
spherical dielectric D2-branes is located at $x^3 = 0$
(Fig.\ref{fig:cyl} (a)).
Since the radius of the spherical D2-brane is given by $h\lambda N$,
adjacent surfaces of the spherical D2-brane 
approach at the length of a string scale at $h\lambda N \sim \ell$. 
There arise a tachyonic mode
in the spectrum of the open strings stretched between those surfaces
(Fig.\ref{fig:cyl} (b)).
This tachyonic mode would imply a deformation of the topology of the sphere 
and, at $h\lambda N > \ell$, the shape of the D2-brane will become a distorted 
torus (Fig.\ref{fig:cyl} (c)).
We call this configuration a torus-like dielectric D2-brane.

The world-volume theory on the D2-brane can be described by the abelian 
Born-Infeld action (\ref{eq:BI}) plus the abelian Chern-Simons 
action (\ref{eq:Che}). We consider the torus-like
D2-brane with $N$ magnetic fluxes in the presence of the constant R-R 4-form
filed strength $G^{(4)}_{0123} = - 4h$, 
thus the relevant fields are the metric $G_{\mu\nu}$, the field strength of 
the gauge field $F_{ab}$ and the R-R 3-form potential $C^{(3)}$.
Since the shape of the D2-brane is a distorted torus, instead of the
static gauge, we employ the gauge
\begin{alignat}{3}
  x^0 = t, \quad x^1 = R(t,z) \cos\theta, \quad x^2 = R(t,z) \sin\theta, \quad
  x^3 =  z.
\end{alignat}
Here $(t,\theta,z)$ are the world-volume coordinates with
$t \in \mathbb{R}$, $-\pi \leq \theta \leq \pi$, and $-\ell \le z \le \ell$.

Now the action of the D2-brane is written in the form
\begin{alignat}{3}
  {\mathcal S}_{\text{D2}} &= - T_2 \int \!\! dt d\theta dz 
  \Big( \sqrt{-\det(P[G]_{ab} + \lambda F_{ab})} \Big) 
  + \mu_2 \int P[C^{(3)}] \notag
  \\
  &= - T_2 \int \!\! dt d\theta dz 
  \bigg\{ \sqrt{ R^2(1+{R'}^2-{\dot{R}}^2) + (\lambda F_{\theta z})^2
  (1-{\dot{R}}^2)} - 2h R^2 \bigg\} , \label{eq:cyla}
\end{alignat}
where $\dot{R}$ and $R'$ are abbreviations of $\frac{dR}{dt}$ and 
$\frac{dR}{dz}$ respectively.
In order to obtain $F_{\theta z}$ as a functional of $R$,
we use the equations of motion for $A_\theta$ and $A_z$. 
These equations lead that the quantity $D$ defined by
\begin{alignat}{3}
  D \equiv - \frac{1}{\lambda T_2}
  \frac{\delta {\mathcal L}}{\delta F_{\theta z}}
  = \frac{\lambda F_{\theta z}(1-\dot{R}^2)}
  {\sqrt{R^2(1+{R'}^2-{\dot{R}}^2) + (\lambda F_{\theta z})^2 
  (1-{\dot{R}}^2)}} ,
\end{alignat}
depends only on $t$. 
By using this $D$, $\lambda F_{\theta z}$ can be expressed as
\begin{alignat}{3}
  \lambda F_{\theta z} = \frac{DR \sqrt{1+{R'}^2-{\dot{R}}^2}}
  {\sqrt{(1-{\dot{R}}^2)(1-{\dot{R}}^2-D^2)}} . \label{eq:F}
\end{alignat}
We only consider the static case hereafter, $\dot{R} = 0$ and $D$ is
a constant. Then the eq.(\ref{eq:F}) becomes simply
\begin{alignat}{3}
  \lambda F_{\theta z} = \frac{D}{\sqrt{(1-D^2)}} R \sqrt{1+{R'}^2} .
  \label{eq:FF}
\end{alignat}
Since the metric on the world-volume is given by $ds^2 = -dt^2 + R^2 d\theta^2
+ (1+{R'}^2)dz^2$, the eq.(\ref{eq:FF}) means that the number of the magnetic 
fluxes per unit area is constant everywhere.
Furthermore, $D$ can be determined by the quantization condition of the
magnetic flux. Now the number of the magnetic fluxes on the D2-brane is $N$,
the condition is written as
\begin{alignat}{3}
  N = \frac{1}{2\pi} \int d\theta dz F_{\theta z}
  = \frac{1}{2\pi\lambda} \frac{D}{\sqrt{(1-D^2)}} S.
\end{alignat}
Here $S$ is the area of the D2-brane world-volume,
\begin{alignat}{3}
  S = \int d\theta dz R\sqrt{1+{R'}^2} .
\end{alignat}
The field strength of the gauge field, therefore, becomes
\begin{alignat}{3}
  \lambda F_{\theta z} = \frac{2\pi\lambda N}{S} R \sqrt{1+{R'}^2} .
\end{alignat}

Using this expression, the action (\ref{eq:cyla}) is 
rewritten in the form
\begin{alignat}{3}
  {\mathcal S}_{\text{D2}} = - T_2 \int \!\! dt 
  \sqrt{S^2 + \big(2\pi\lambda N \big)^2} + 4\pi h T_2 \int \!\! dt dz R^2 .
\end{alignat}
and the potential energy is given by
\begin{alignat}{3}
  V_{\text{D2}} = \sqrt{\big(S T_2 \big)^2 + \big(NT_0 \big)^2}
  - 4\pi h T_2 \int \!\! dz R^2 . \label{eq:pot3}
\end{alignat}
The first term is the square root of the sum of the D2-brane mass squared and
$N$ D0-branes mass squared, as expected.

Now the potential energy is obtained as a functional of $R(z)$.
The next task is to get the explicit form of $R(z)$.
The equation of motion for $R$ is obtained by varing the potential energy.
\begin{alignat}{3}
  \frac{\delta V_{\text{D2}}}{\delta R} &=
  \frac{\partial V_{\text{D2}}}{\partial R} - 
  \Big(\frac{\partial V_{\text{D2}}}{\partial R'} \Big)' \notag
  \\[3pt]
  &= \frac{\partial V_{\text{D2}}}{\partial S} \frac{\partial S}{\partial R}
  - 8\pi h T_2 R - \Big( \frac{\partial V_{\text{D2}}}{\partial S}
  \frac{\partial S}{\partial R'} \Big)' \notag
  \\
  &= \frac{2\pi T_2}{R'} \bigg\{ \frac{R}{P\sqrt{1+{R'}^2}} - 2hR^2 \bigg\}' 
  \label{eq:vpot}
  \\
  &= 0 . \notag
\end{alignat}
Here $P$ is defined as
\begin{alignat}{3}
  P = \sqrt{1 + \Big(\frac{NT_0}{ST_2} \Big)^2} .
\end{alignat}
One can easily integrate the eq.(\ref{eq:vpot}) once and obtain
\begin{alignat}{3}
  R' &= \pm \frac{2h}{2hR^2 \!+\! C} \sqrt{(R_+^2 - R^2)(R^2 - R_-^2)} ,
  \label{eq:R'}
\end{alignat}
where $C$ is an integration constant and $R_\pm^2$ are defined as
\begin{alignat}{3}
  R_{\pm}^2 &\equiv \frac{C}{hX} \big( 1 - X/2 \pm \sqrt{1 - X} \big) ,
  \label{eq:X}
  \\[3pt]
  X & \equiv 8ChP^2 . \notag
\end{alignat}
If $X \le 1$, the eq.(\ref{eq:R'}) shows that $R'\!=\!0$ at $R\!=\!R_\pm$ 
and $R$ is periodic with $R_- \!\le\! R \!\le\! R_+$. 
Because we are studying the configurations like Fig.\ref{fig:cyl}(c),
the minus (plus) sign of the eq.(\ref{eq:R'}) is chosen 
in $0 \!\le\! z \!\le\! \ell$ ($-\ell \!\le\! z \!\le\! 0$). 
We only consider the region where $0 \!\le\! z \!\le\! \ell$ in the following.
The eq.(\ref{eq:R'}) is further integrated as
\begin{alignat}{3}
  z &= \frac{1}{2h} \int_R^{R_+} \!\!\!\! dR
  \frac{2hR^2 \!+\! C}{\sqrt{(R_+^2 - R^2)(R^2 - R_-^2)}} \notag
  \\[3pt]
  &= R_+ \int_0^{\psi} \!\! d\phi \sqrt{1-k^2 \sin^2 \phi}
  + \frac{C}{2h R_+} \int_0^{\psi} \!\! d\phi 
  \frac{1}{\sqrt{1-k^2 \sin^2 \phi}} \label{eq:z}
  \\[-3pt]
  &= R_+ E(\psi,k) + \frac{C}{2h R_+} F(\psi,k) , \notag
\end{alignat}
where we choose such an integration constant that $R = R_+$ at $z=0$ and
$R = R_-$ at $z=\ell$.
$F(\psi,k)$ and $E(\psi,k)$ are the incomplete elliptic integrals of
the first and second kinds respectively, $k$ is the elliptic modulus 
and $\psi$ is the angle given by
\begin{alignat}{3}
  k &= \frac{\sqrt{R_+^2-R_-^2}}{R_+} ,\quad
  \psi = \sin^{-1} \sqrt{\frac{R_+^2 - R^2}{R_+^2 - R_-^2}} . \label{eq:k}
\end{alignat}
The range of the elliptic modulus is $0 \!\le\! k \!\le\! 1$ and
that of the angle is $0 \!\le\! \psi \!\le\! \frac{\pi}{2}$.
Some properties of the elliptic integrals are listed in the
appendix \ref{sec:elli}.

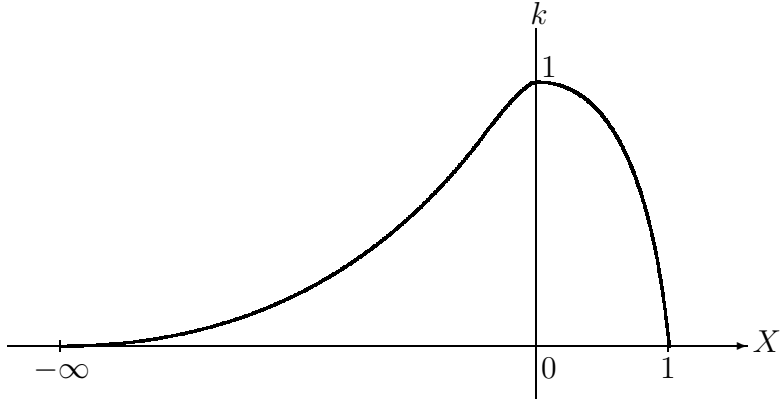
\begin{figure}[tb]
\begin{center}
  \begin{picture}(300,150)
    \put(0,20){\vector(1,0){280}}
    \put(200,0){\line(0,1){140}}
    \put(282,18){$X$}
    \put(198,142){$k$}
    \put(202,8){0}                
    \put(250,18){\line(0,1){4}}    
    \put(247,8){$1$}
    \put(198,120){\line(1,0){4}}   
    \put(202,122){$1$}
    \put(20,18){\line(0,1){4}}     
    \put(10,8){$-\infty$}
    \thicklines
    \qbezier(20,20)(120,20)(180,100)
    \qbezier(180,100)(195,120)(200,120)
    \qbezier(200,120)(240,120)(250,20)
  \end{picture}
  \caption{$k$ as a function of $X$.}
  \label{fig:k}
\end{center}
\end{figure}

Now we will examine the solutions (\ref{eq:z}) in detail.
For this purpose, it is better to express the quantities in terms of
the modulus $k$. As we will see later, however, two distinct solutions 
correspond to the same $k$. 
In fact, $X \!=\! 8ChP^2$ is more suitable for a parameter of solutions.
By the definition (\ref{eq:k}), $k$ can be expressed as a function of $X$.
\begin{alignat}{3}
  k^2 = \frac{2 \sqrt{1 - X} }{1 - X/2 + \sqrt{1 - X} }.
\end{alignat}
The range of $X$ is restricted to $X \!\le\! 1$, which is consistent with
$0 \!\le\! k \!\le\! 1$. In general, two distinct
values of $X$ correspond to the same $k$, for example, 
$k\!=\!0$ at $X\!=\!-\infty,1$. The only exception is $k\!=\!1$ at $X\!=\!0$.
$k$ increase monotonously in the region of $X \!<\! 0$, and 
decrease monotonously in that of $0 \!<\! X \!<\! 1$ (Fig.\ref{fig:k}).

Now we rewrite the quantities obtained so far as functions of $X$.
By the definition of $S$, we obtain
\begin{alignat}{3}
  S = \frac{2\pi}{hP} \int_{R_-}^{R_+} \!\!\!\! dR
  \frac{R^2}{\sqrt{(R_+^2 - R^2)(R^2 - R_-^2)}} 
  = \frac{2\pi R_+}{hP} E(k), \label{eq:area}
\end{alignat}
where $E(k)$ is the complete elliptic integral of the second kind.
By using this equation (\ref{eq:area}), $P$ is written as
\begin{alignat}{3}
  P = \frac{1}{\sqrt{1 - \bigg( \dfrac{h\lambda N}{R_+ E(k)} \bigg)^2}}.
  \label{eq:P}
\end{alignat}
By using this equation (\ref{eq:P}) and the definition of $X$ (\ref{eq:X}),
we can rewrite the constant $C$ by means of $X$ as
\begin{alignat}{3}
  C &= \frac{X}{16h} \left( 1 - \sqrt{1 - \frac{32(h^2\lambda N)^2}
  {E(k)^2 \big(1 \!-\! X/2 \!+\! \sqrt{1 \!-\! X}\big)} } \, \right) \notag
  \\
  &\sim \frac{X(h^2\lambda N)^2}{h E(k)^2 
  \big(1 \!-\! X/2 \!+\! \sqrt{1 \!-\! X}\big)} ,
\end{alignat}
where $\sim$ holds when $h^2\lambda N \ll 1$.
Here we take a choice which gives the smaller area.
$S$ is now written in the form
\begin{alignat}{3}
  S &= 4\pi(h\lambda N)^2 \times \frac{2\sqrt{2}}
  {E(k) \sqrt{1 \!-\! X/2 \!+\! \sqrt{1 \!-\! X}} }
  \left( 1 + \sqrt{1 - \frac{32(h^2\lambda N)^2}
  {E(k)^2 \big(1 \!-\! X/2 \!+\! \sqrt{1 \!-\! X}\big)} } \, \right)^{-1} 
  \notag
  \\
  &\sim 4\pi(h\lambda N)^2 \times \frac{\sqrt{2}}
  {E(k) \sqrt{1 \!-\! X/2 \!+\! \sqrt{1 \!-\! X}} }.
\end{alignat}
In the limit of $h^2\lambda N \ll 1$, 
$S$ is equal to $4\pi (h\lambda N)^2$ at $X\!=\!0$. The solution of
$X\!=\!0$, therefore, corresponds to the spherical D2-brane 
in the previous section.

$R_\pm$ are given by
\begin{alignat}{3}
  R_\pm^2 &= \frac{1 \!-\! X/2 \!\pm\! \sqrt{1\!-\!X} }{16h^2}
  \left( 1 - \sqrt{1 - \frac{32(h^2\lambda N)^2}
  {E(k)^2 \big(1 \!-\! X/2 \!+\! \sqrt{1 \!-\! X}\big)} } \, \right), \notag
  \\
  R_+ &\sim \frac{h\lambda N}{E(k)},
  \\
  R_- &\sim \frac{h\lambda N}{E(k)} \times \sqrt{
  \frac{ 1 \!-\! X/2 \!-\! \sqrt{1 \!-\! X} }
  { 1 \!-\! X/2 \!+\! \sqrt{1 \!-\! X} } } .
\end{alignat}
$R_+ \!=\! R_-$ when $X \!=\! 1$, thus the shape of the D2-brane becomes
a torus. Up to order $h^2\lambda N$, 
the radius of the cycle, which does not wind any circle of the target space, 
equals to $\frac{2}{\pi}h\lambda N$.

The half of the circumference of the $x^3$ circle $\ell$ is given by
\begin{alignat}{3}
  \ell &= R_+ E(k) + \frac{C}{2hR_+}K(k) \notag
  \\
  &= \frac{1}{4h} \!\! \left( E(k) \sqrt{1 \!\!-\!\! X/2 \!\!+\!\! 
  \sqrt{1\!\!-\!\!X}} + \frac{K(k)X}{2\sqrt{1 \!\!-\!\! X/2 \!\!+\!\! 
  \sqrt{1\!\!-\!\!X}} }\right) \!\!
  \sqrt{ 1 \!-\! \sqrt{1 \!-\! \frac{32(h^2\lambda N)^2}
  {E(k)^2 \big(1 \!\!-\!\! X/2 \!\!+\!\! \sqrt{1 \!\!-\!\! X}\big)} } \, }
  \notag
  \\
  &\sim h\lambda N \times \left( 1 + \frac{K(k) X}{2E(k)
  \big(1 \!\!-\!\! X/2 \!\!+\!\! \sqrt{1 \!\!-\!\! X}\big)} \right) .
  \label{eq:ell}
\end{alignat}
$K(k)$ is the complete elliptic integral of the first kind.
Notice that $\ell$ depends on two quantities, $h\lambda N$ and $X$, 
in the limit of $h^2 \lambda N \ll 1$.
As it is usual to fix $\ell$, the eq.(\ref{eq:ell}) is interpreted as a
relation between $h\lambda N$ and $X$.
$X$ and $h\lambda N$ govern the shape and size
of the D2-brane respectively.

Finally $z$ is written as
\begin{alignat}{3}
  z &= \frac{1}{4h} \!\! \left( E(\psi,k) \sqrt{1 \!\!-\!\! X/2 \!\!+\!\! 
  \sqrt{1\!\!-\!\!X}} + \frac{F(\psi,k)X}{2\sqrt{1 \!\!-\!\! X/2 \!\!+\!\! 
  \sqrt{1\!\!-\!\!X}} }\right) \!\!
  \sqrt{ 1 \!-\! \sqrt{1 \!-\! \frac{32(h^2\lambda N)^2}
  {E(k)^2 \big(1 \!\!-\!\! X/2 \!\!+\!\! \sqrt{1 \!\!-\!\! X}\big)} } \, }
  \notag
  \\
  &\sim h\lambda N \times \left( \frac{E(\psi,k)}{E(k)} 
  + \frac{F(\psi,k) X}{2E(k)
  \big(1 \!\!-\!\! X/2 \!\!+\!\! \sqrt{1 \!\!-\!\! X}\big)} \right) ,
\end{alignat}
and the potential energy is given by
\begin{alignat}{3}
  V_{\text{D2}} &= \sqrt{\big(S T_2 \big)^2 + \big(NT_0 \big)^2}
  - 4\pi h T_2 \left\{ \frac{4(2-k^2)E(k)}{3} R_+^3 - \frac{2(1-k^2)K(k)}{3}
  R_+^3 + E(k) \frac{CR_+}{h} \right\} \notag
  \\
  &\sim NT_0 - NT_0 \left\{ \frac{8(2-k^2)}{3E(k)^2} - 
  \frac{4(1-k^2)K(k)}{3E(k)^3} -
  \frac{4-2X}{E(k)^2\big(1 \!\!-\!\! X/2 \!\!+\!\! \sqrt{1 \!\!-\!\! X}\big)} 
  \right\} (h^2\lambda N)^2 . \label{eq:ene}
\end{alignat}
The results in the limit of $h^2\lambda N \ll 1$ 
are collected in Table \ref{tab:sum}.
As already mentioned, we see that $X\!=\!0$ corresponds to the spherical 
D2-brane which agrees with the one obtained in \S\S\ref{subsec:sph}
and $X\!=\!1$ corresponds to the toroidal D2-brane
whose potential energy is equal to that of $N$ D0-branes.
Even though the potential energy takes minimum for the spherical D2-brane,
if $h^2\lambda N$ is small enough we could expect that the 
torus-like D2-branes are meta-stable.

\begin{figure}[htb]
  \begin{center}
    \includegraphics[width=8cm,height=2cm]{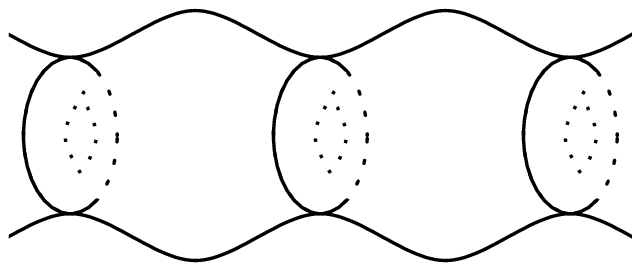}
  \begin{picture}(0,0)
    \put(-55,-2){$z$}
    \put(-120,0){\vector(1,0){60}}
  \end{picture}
  \caption{The case of $X < 0$.} 
  \label{fig:tortor}
  \end{center}
\end{figure}

\begin{figure}[htb]
  \begin{center}
    \includegraphics[width=5cm,height=8cm,keepaspectratio]{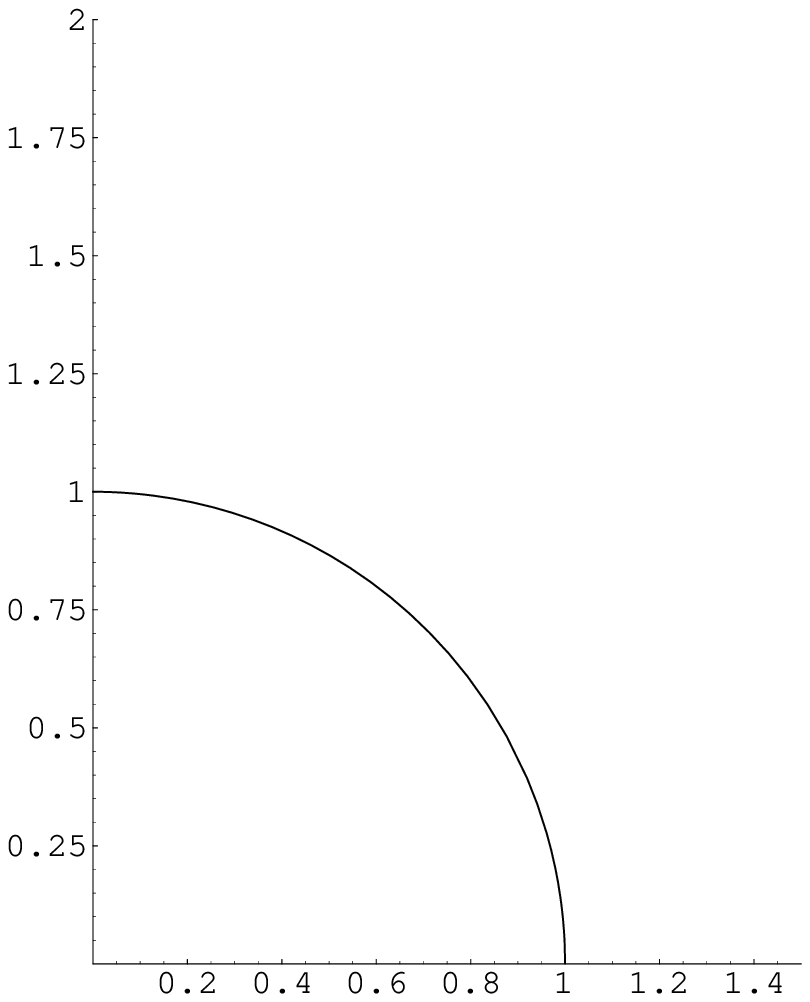}
    \includegraphics[width=5cm,height=8cm,keepaspectratio]{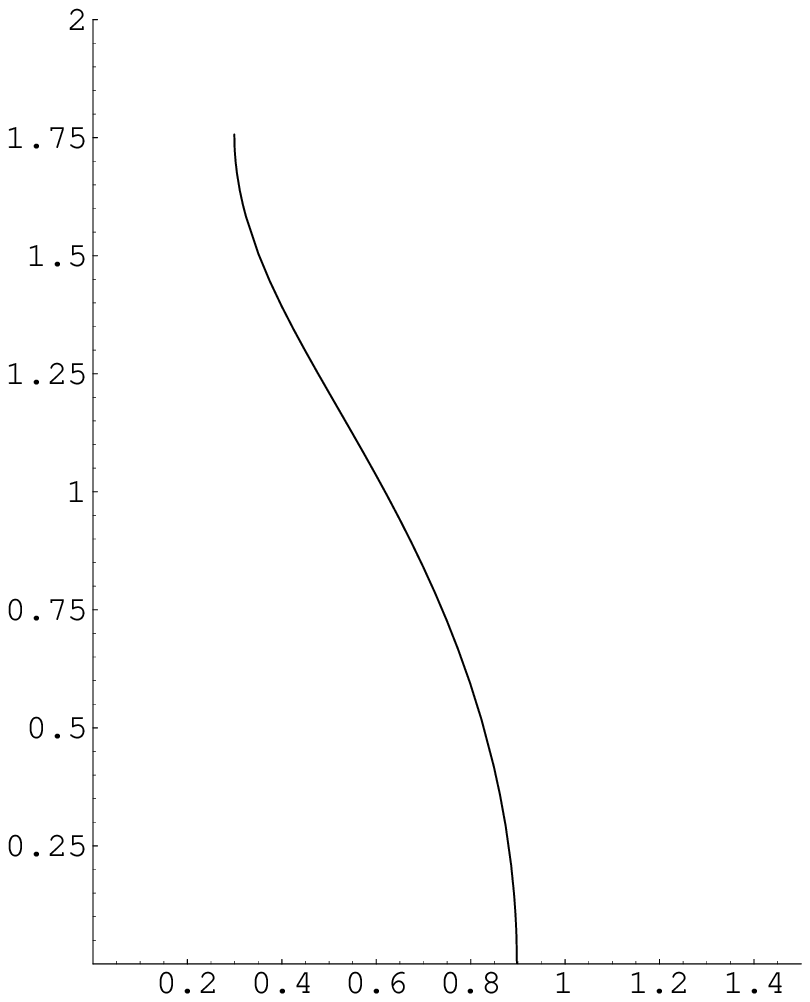}
    \includegraphics[width=5cm,height=8cm,keepaspectratio]{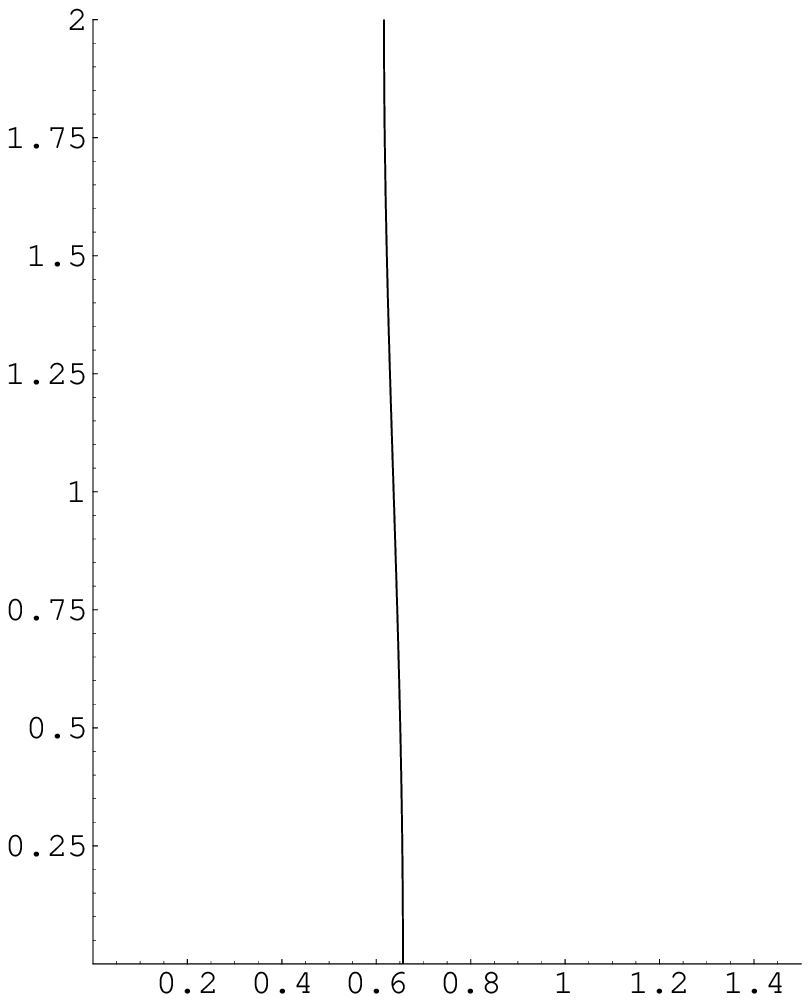}
    \\[1cm]
    \includegraphics[width=5cm,height=8cm,keepaspectratio]{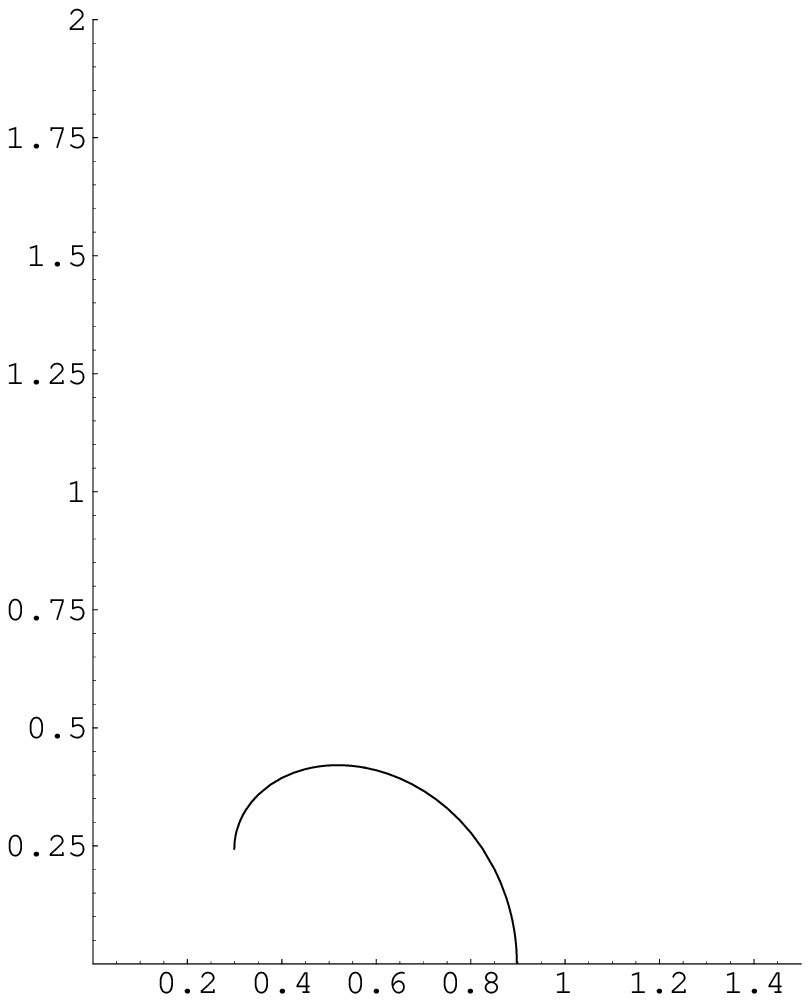}
    \includegraphics[width=5cm,height=8cm,keepaspectratio]{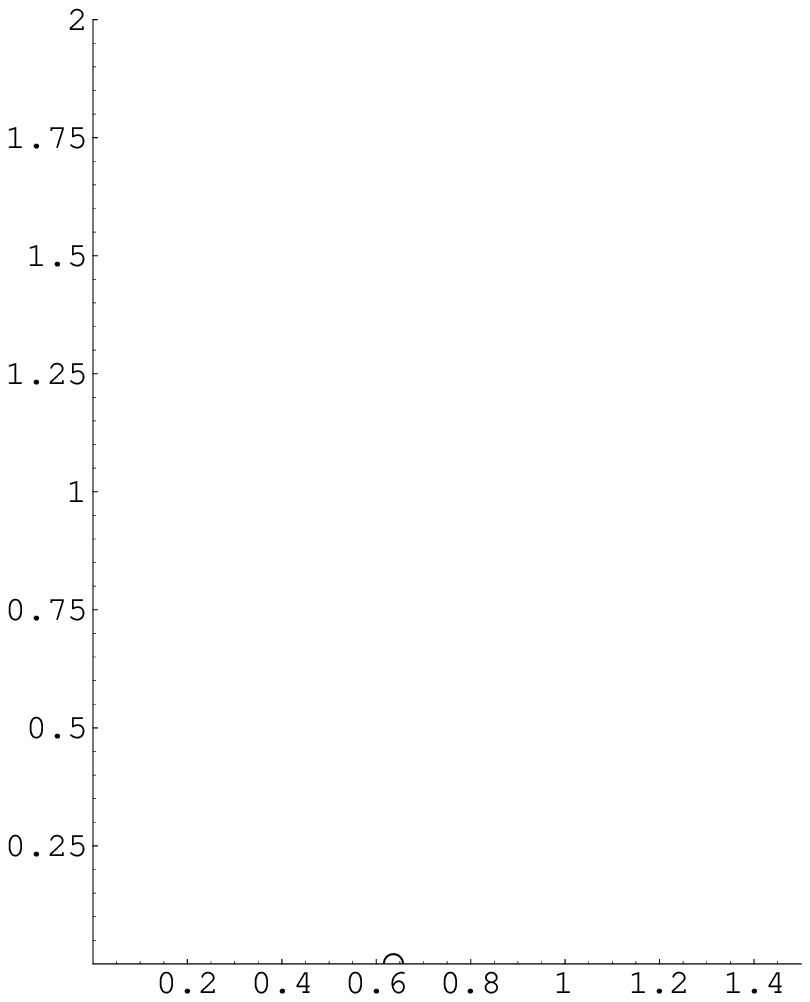}
  \begin{picture}(450,10)
    \put(70,300){$X=0$}
    \put(230,300){$X=0.75$}
    \put(375,300){$X=0.999$}
    \put(150,60){$X=-3$}
    \put(290,60){$X=-999$}
    \put(80,205){(a)}
    \put(225,205){(b)}
    \put(370,205){(c)}
    \put(150,0){(d)}
    \put(300,0){(e)}
  \end{picture}
  \caption{The shapes of D2-branes at $X \!\!=\!\! 0$, $0.75$, $0.999$,
  $-3$ and $-999$.
  Vertical axes label $z/h\lambda N$ and horizontal axes label
  $R/h\lambda N$. $R_-/h\lambda N$ of (a), (b), (c), (d) and (e) are equal to
  $0$, $0.299$, $0.616$, $0.299$ and $0.616$ respectively.}
  \label{fig:D2}
  \end{center}
\end{figure}

The shapes of the torus-like D2-branes where $X \!=\! 0$, $0.75$, $0.999$
$-3$, $-999$ are depicted in Fig.\ref{fig:D2}. 
For these solutions, $R_-/h\lambda N = 0$, $0.299$, $0.616$, $0.299$ ,$0.616$, 
respectively. The remarkable D2-brane configuration is like 
Fig.\ref{fig:D2}(d). In this case, another torus-like D2-brane seems to appear
in the interior of the original torus-like D2-brane(Fig.\ref{fig:tortor}).
However, the world-volume of the D2-brane intersect at $z=\ell$.
The space-metric would not be flat around there and our 
calculations might go wrong. For this reason, we do not take seriously
the results for the region $-\infty < X < 0$.

\vspace{1cm}
\begin{table}[htb]
  \begin{center}
  \renewcommand{\arraystretch}{1.4}
  \setlength{\tabcolsep}{14pt}
  \begin{tabular}{c|c|c|c|c|c}
    $X$  &  $-\infty$ & $\cdots$ &  $0$ (sphere)  
    & $\cdots$ &  $1$ (torus)  \\ \hline
    $\ell$  &  $0$  &  $\nearrow$  &  $h\lambda N$  
    &  $\nearrow$  &  $2h\lambda N$  \\ \hline
    $R_+$  &  $\frac{2}{\pi}h\lambda N$  &  $\nearrow$  &  $h\lambda N$  
    &  $\searrow$  &  $\frac{2}{\pi}h\lambda N$  \\ \hline
    $R_-$  &  $\frac{2}{\pi}h\lambda N$  &  $\searrow$  &  $0$  
    &  $\nearrow$  &  $\frac{2}{\pi}h\lambda N$  \\ \hline
    $V_{\text{D2}}$  &  $NT_0$  &  $\searrow$  &  
    $NT_0 - \frac{2}{3} NT_0 (h^2\lambda N)^2$  
    &  $\nearrow$  &  $NT_0$  \\
  \end{tabular}
  \caption{The behaviors of $\ell$, $R_\pm$ and $V_{\text{D2}}$ in the 
  limit of $h^2\lambda N \ll 1$.}
  \label{tab:sum}
  \end{center}
\end{table}

\newpage

\vspace{1cm}
\section{Torus-like Generalization of Myers Effect}
\label{sec:TorMye}

In the previous section, we obtained the various torus-like configurations
of the D2-brane, including the spherical configuration.
As we have seen in  the \S\S\ref{subsec:Mye}, the spherical dielectric
D2-brane could also be obtained from the world-volume theory of the
$N$ D0-branes. In this section, we express the torus-like D2-branes
from the viewpoint of the world-volume theory of the $N$ D0-branes.

We consider the same system as in the \S\S\ref{subsec:Mye}. The potential
energy and the equation of motion are given by the eq.(\ref{eq:pot2}) and 
the eq.(\ref{eq:eom2}) respectively. 
Our aim is to find the torus-like solutions of the eq.(\ref{eq:eom2}).
Let us start the explicit matrix forms of the
spherical solution (\ref{eq:J_3}),(\ref{eq:J_+-}).
A small deformation of the spherical configuration would not change the
positions of non-zero components in the matrices so much.
Therefore, we assume the form of $J_3$ as
\begin{alignat}{3}
  J_3 &= 
  \begin{pmatrix}
    \; \dd & & & & & & & & \;\\
    \: & z_{N} \pp 2\ell' & & & & & & & \;\\
    \; & & z_1 & & & & & & \;\\
    \; & & & z_2 & & & & & \;\\
    \: & & & & \dd & & & & \;\\
    \; & & & & & z_{N-1} & & & \;\\
    \; & & & & & & z_{N} & & \;\\
    \: & & & & & & & z_1 \mm 2\ell' & \;\\
    \; & & & & & & & & \dd \;
  \end{pmatrix} . \label{eq:J3}
\end{alignat}
Note that the compactification of the $x^3$ direction makes the size of
matrix infinity. The $N\times N$ diagonal block containing the 
components $z_1,\cdots,z_N$ are the original matrix and the others
are the copies of it.
$z_1,\cdots,z_N$ multiplied by $2h\lambda$ denote the positions 
in the $x^3$ direction 
of $N$ D0-branes and $2\ell'$ multiplied by $2h\lambda$ 
is the circumference of the compactified circle. We also assume the forms of 
$J_\pm$ as
\begin{alignat}{3}
  J_+ = J_-^\dagger &=
  \begin{pmatrix}
    \; \dd\!\! & R_{N \m 1} & & & & & & & \;\\
    \: & 0 & R_N & & & & & & \;\\
    \; & & 0 & \,\;R_1\; & & & & & \;\\
    \; & & & 0 & & & & & \;\\
    \: & & & & \;\dd\; & & & & \;\\
    \; & & & & & 0 & R_{N \m 1} & & \;\\
    \; & & & & & & 0 & R_N & \;\\
    \: & & & & & & & 0 & \;R_1\; \;\\
    \; & & & & & & & & \dd \;
  \end{pmatrix} . \label{eq:J+-}
\end{alignat}
$R_1,\cdots,R_N$ multiplied by $2h\lambda$ 
represent the extension, to the radial direction in the $x_1$-$x_2$ plane, 
of the cell between adjacent D0-branes.
Because the configurations of D0-branes should be symmetric with respect to 
the reversal at $z=0$, the conditions $R_n \!=\! R_{N\m n}$
$(n=1,\cdots,[\frac{N-1}{2}])$ must be imposed.

These assumptions are justified by comparing the
potential energy (\ref{eq:pot2}) and (\ref{eq:pot3}).
By substituting the assumptions (\ref{eq:J3}),(\ref{eq:J+-}) 
for the eq.(\ref{eq:pot2}),  
we obtain for the original $N\times N$ matrix
\begin{alignat}{3}
  V_{\text{$N$D0}} &= NT_0 + \frac{8NT_0 (h^2\lambda N)^2}{N^3}
  \sum_{n=1}^N (z_n \mm z_{n\p 1})^2 \bigg\{ R_n^2
  + \frac{1}{4} \bigg(\frac{ R_n^2 \mm R_{n\p 1}^2 }{z_n \mm z_{n\p 1}} 
  \bigg)^2 \bigg\} \notag
  \\
  & \qquad\quad\; - \frac{16NT_0 (h^2\lambda N)^2}{N^3}
  \sum_{n=1}^N R_n^2 (z_n \mm z_{n\p 1}) \; . \label{eq:VND0}
\end{alignat}
On the other hand, by using the approximations
\begin{alignat}{3}
  \int dz R^2 &\sim (2h\lambda)^3 \sum_{i=1}^N R_n^2 (z_n \mm z_{n\p 1}) \;,
  \\
  \int dz R \sqrt{1 + {R'}^2} &= \int dz \sqrt{R^2 + \tfrac{1}{4} \{(R^2)'\}^2}
  \sim (2h\lambda)^2 \sum_{n=1}^N (z_n \mm z_{n\p 1}) \sqrt{ R_n^2
  + \frac{1}{4} \bigg(\frac{ R_n^2 \mm R_{n\p 1}^2 }{z_n \mm z_{n\p 1}} 
  \bigg)^2 } \;, \notag
\end{alignat}
the eq.(\ref{eq:pot3}) is evaluated in this case as
\begin{alignat}{3}
  V_{\text{D2}} &\sim NT_0 + \frac{1}{2} \left( \frac{ST_2}{NT_0} \right)^2
  - 4\pi h T_2 \int dz R^2 \notag
  \\
  &= NT_0 + \frac{8NT_0 (h^2\lambda N)^2}{N^4}
  \Bigg\{ \sum_{n=1}^N (z_n \mm z_{n\p 1}) \sqrt{ R_n^2
  + \frac{1}{4} \bigg(\frac{ R_n^2 \mm R_{n\p 1}^2 }{z_n \mm z_{n\p 1}} 
  \bigg)^2 } \; \Bigg\}^2 \label{eq:VD2}
  \\
  & \qquad\quad\; - \frac{16NT_0 (h^2\lambda N)^2}{N^3}
  \sum_{n=1}^N R_n^2 (z_n \mm z_{n\p 1}) \; . \notag
\end{alignat}
The second terms of two potential energies (\ref{eq:VND0}),(\ref{eq:VD2}) 
are different. However, if the conditions,
\begin{alignat}{3}
  (z_n \mm z_{n\p 1}) \sqrt{ R_n^2 + \frac{1}{4} 
  \bigg(\frac{ R_n^2 \mm R_{n\p 1}^2 }{z_n \mm z_{n\p 1}} \bigg)^2 }
  = K N + \mathcal{O}(1) , \label{eq:ass}
\end{alignat}
hold for some constant $K$ and arbitrary $n$, two potential energies
become equal up to the order $\frac{1}{N}$. 
The left hand side of the eq.(\ref{eq:ass}) is regarded as 
the measure $dz R\sqrt{1\p {R'}^2}$ in the large $N$ limit, and thus the 
eq.(\ref{eq:ass}) implies that the areas of cells between the adjacent 
D0-branes are a constant. The solutions of the eq.(\ref{eq:eom2}) might 
satisfy these conditions. 
For example, $K$ is equal to $\frac{1}{2}$ for the fuzzy two-sphere.

By substituting the assumptions (\ref{eq:J3}),(\ref{eq:J+-}) 
for the eq.(\ref{eq:eom2}), we obtain equations
\begin{alignat}{3}
  &R_n \left\{ (z_n \m z_{n\p 1})^2 - 2 (z_n \m z_{n\p 1}) + R_n^2 - 
  \frac{R_{n\m 1}^2 \p R_{n\p 1}^2}{2} \right\} = 0 , \notag
  \\
  &- R_n^2 (z_n \m z_{n\p 1} \m 1) + R_{n\m 1}^2 (z_{n\m 1} \m z_n \m 1) = 0 ,
  \\[1pt]
  &\qquad\qquad\qquad n = 1,\cdots,N, \notag
\end{alignat}
where $R_0 \!=\! R_N$, $R_{N\p 1} \!=\! R_1$, 
$z_0 \!=\! z_N \p 2\ell'$, $z_{N\p 1} \!=\! z_1 \m 2\ell'$ 
and so on. Further calculation gives
\begin{alignat}{3}
  &z_n \m z_{n\p 1} = 1 \pm \sqrt{1 - R_n^2 +
  \frac{R_{n\m 1}^2 \p R_{n\p 1}^2}{2}} , \notag
  \\
  &R_n^2 \sqrt{1 - R_n^2 + \frac{R_{n\m 1}^2 \p R_{n\p 1}^2}{2}}
  = R_{n\m 1}^2 \sqrt{1 - R_{n\m 1}^2 + \frac{R_{n\m 2}^2 \p R_n^2}{2}} ,
  \\[1pt]
  &\qquad\qquad\qquad\qquad n=1,\cdots,N . \notag
\end{alignat}
It is difficult to solve these equations explicitly. In principle, however, 
if we fix $R_N$, all the other $R_n$ and $(z_n \m z_{n\p 1})$ are 
written in terms of $R_N$. One of the solutions is $R_1= \cdots = R_N$ and
this corresponds to the toroidal configuration.
Another non-trivial one is
\begin{alignat}{3}
  R_N = 0 \;,\; R_n = \sqrt{n(N-n)} \;,\; 
  z_0 = \frac{N\p 1}{2} \pm \sqrt{N} \;,\;z_n = \frac{N\p 1 \m 2n}{2} ,
\end{alignat}
where $n = 1,\cdots,N\mm 1$. If $N$ is large, this solution corresponds to
the fuzzy two-sphere.

\begin{figure}[htb]
  \begin{center}
    \includegraphics[width=5cm,height=8cm,keepaspectratio]{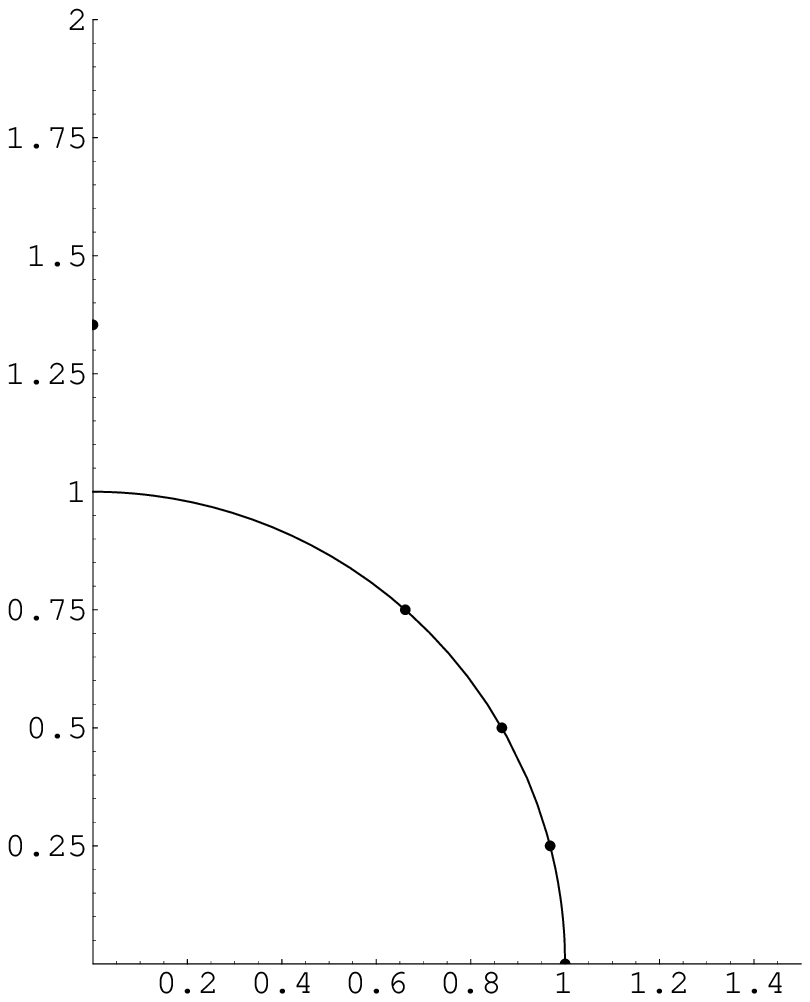}
    \includegraphics[width=5cm,height=8cm,keepaspectratio]{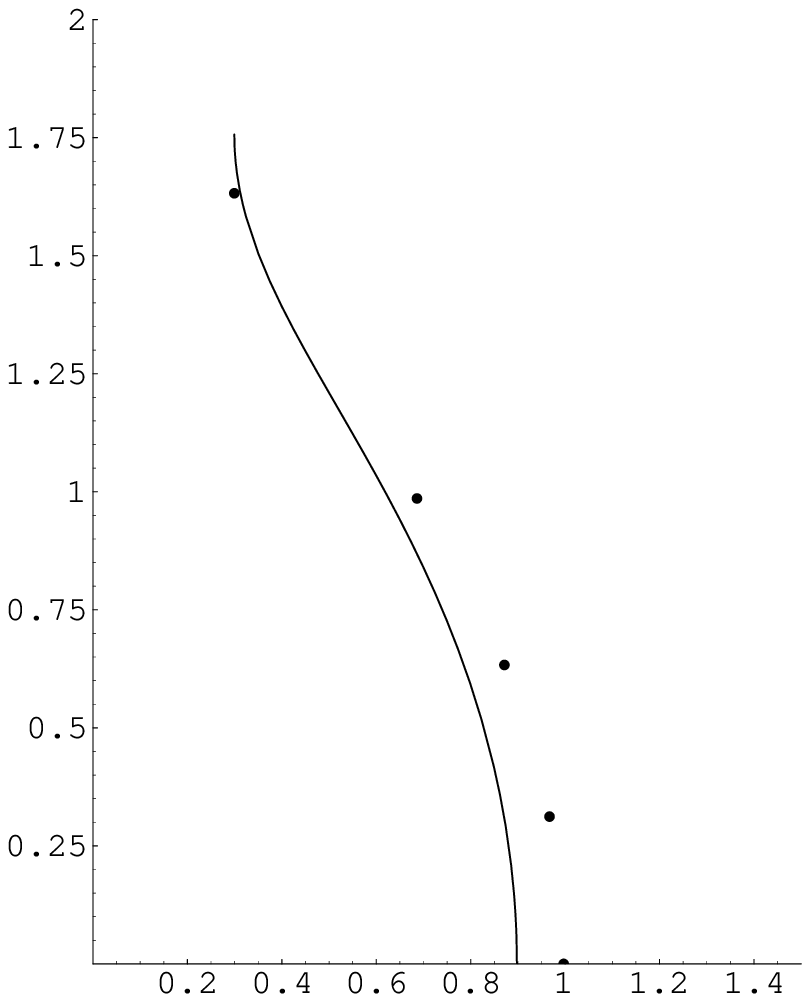}
    \includegraphics[width=5cm,height=8cm,keepaspectratio]{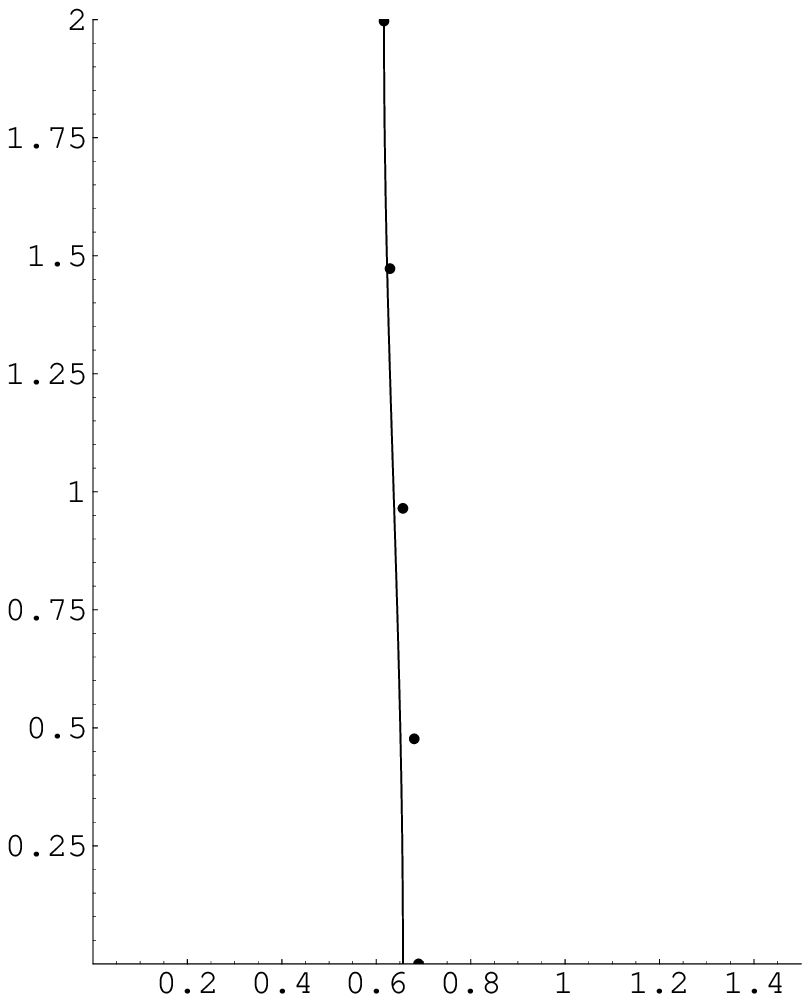}
    \\[1cm]
    \includegraphics[width=5cm,height=8cm,keepaspectratio]{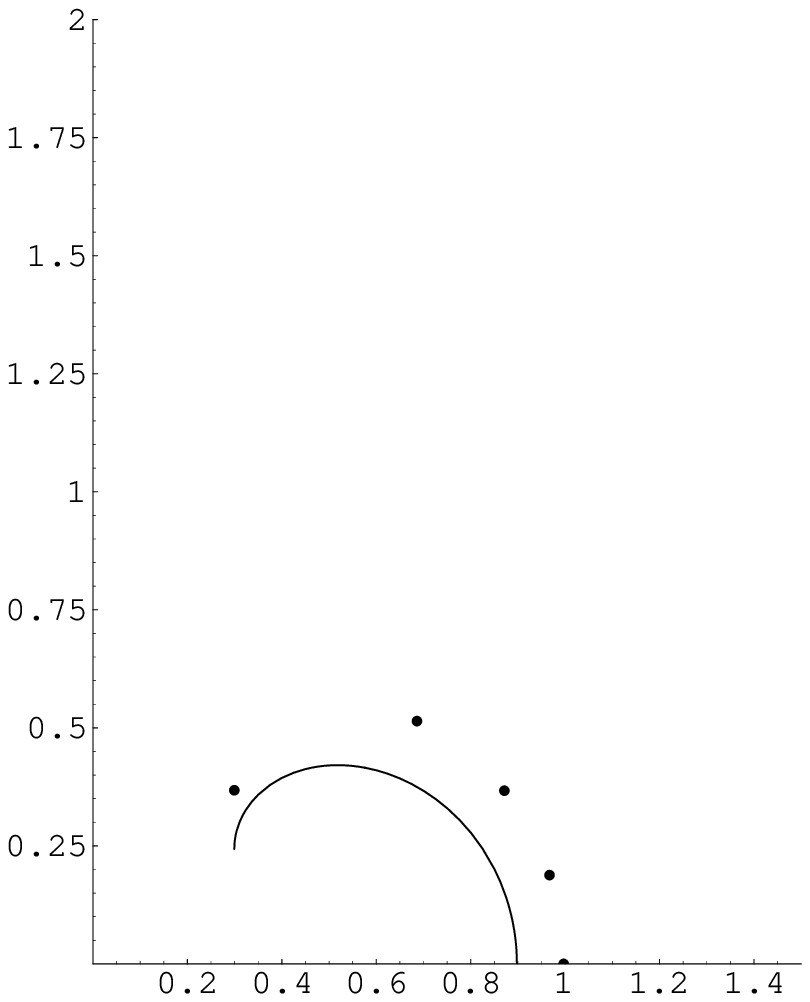}
    \includegraphics[width=5cm,height=8cm,keepaspectratio]{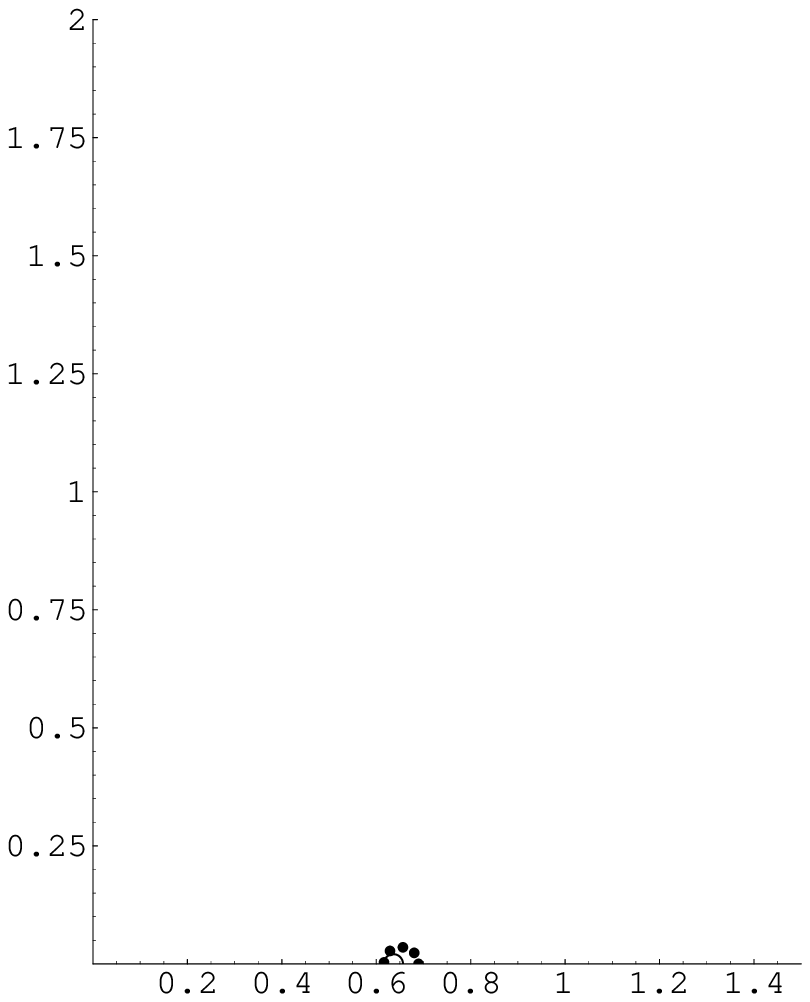}
  \begin{picture}(450,10)
    \put(80,205){(a)}
    \put(225,205){(b)}
    \put(370,205){(c)}
    \put(150,0){(d)}
    \put(300,0){(e)}
  \end{picture}
  \caption{Vertical axes label $z/h\lambda N$ and horizontal axes label
  $R/h\lambda N$. Solid lines are same as the Fig.\ref{fig:D2}. 
  In each figure, dots represent the points, 
  $(\frac{R_4}{4},0)$, $(\frac{R_3}{4},\frac{z_3+z_4}{8})$, 
  $(\frac{R_2}{4},\frac{z_2+z_3}{8})$, $(\frac{R_1}{4},\frac{z_1+z_2}{8})$ 
  and $(\frac{R_8}{4},\frac{z_0+z_1}{8})$. }
  \label{fig:8D0}
  \end{center}
\end{figure}

In order to get some insights, we explicitly examine $N=3$ case here.
In this case, $R_1$ is equal to $R_2$ and given as a function of $R_3$,
\begin{alignat}{3}
  R_1 = \sqrt{1 + \sqrt{2 \Big(\frac{\sqrt{3}+1}{2}-R_3^2 \Big)
  \Big( R_3^2 + \frac{\sqrt{3}-1}{2} \Big)} } .
\end{alignat}
In the range of $0 \le R_3 \le \frac{1}{\sqrt{2}}$, $R_1$ becomes larger as
$R_3$ increase. In $\frac{1}{\sqrt{2}} \le R_3 \le \frac{4}{3}$,
$R_1$ becomes smaller as $R_3$ increase. $R_1 = \sqrt{2}$ at $R_3 = 0, 1$ and
$R_1 = R_3$ at $R_3 = \frac{4}{3}$.
We should mention that $R_1$ does not take the maximum when $R_3$ 
takes the minimum. The results of numerical calculations for $N=8$ are
given in Fig.\ref{fig:8D0}.

\newpage

\vspace{1cm}
\section{Discussions}
\label{sec:Tac}

In this paper, in the presence of R-R flux $G^{(4)}_{0123} = -4h$, 
we have obtained, by compactifing the $x^3$ direction with
the circumference $2\ell$, solutions of torus-like dielectric 
D2-brane with $N$ magnetic fluxes. 
The analyses were done in two different ways.
From the viewpoint of the world-volume action of the D2-brane, we have
seen that, in the limit of $h^2\lambda N \ll 1$, 
the shape and size of the torus-like D2-brane are 
governed by $X (0 < X < 1)$ and $h\lambda N$ respectively.  
$X=0$ corresponds to the spherical configuration with the radius
$h\lambda N$ and $X=1$ does to the toroidal one whose radius of the 
cycle, which does not wind any circle of the target space, 
is equal to $\frac{2}{\pi}h\lambda N$.
The potential energy of solutions increases as $X$ does.
Thus the spherical configuration is the stablest and the others
are meta-stable. 

From the viewpoint of the world-volume action of the $N$ D0-branes,
we have put the assumptions for the components of the matrices 
$J_3$, $J_\pm$ and obtained the relations between them.
These relations can be solved in principle. 
The results of the numerical calculations were given in the case 
of $N=8$ and they match well to the solutions obtained from
the D2-brane action.

\begin{figure}[htb]
  \begin{center}
    \includegraphics[width=12cm,height=6cm,keepaspectratio]{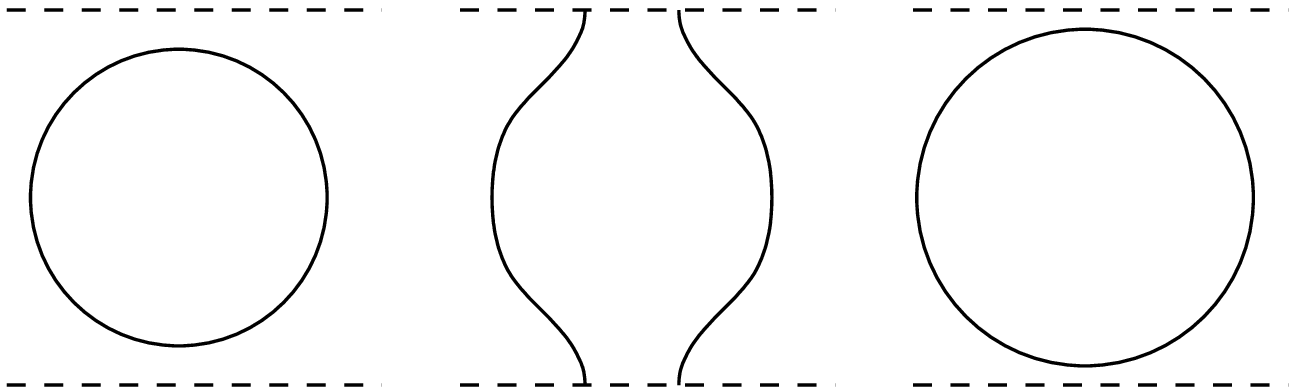}
  \begin{picture}(400,10)
    \put(7,128){$z$}
    \put(10,5){\line(0,1){120}}
    \put(8,17){\line(1,0){4}}
    \put(-9,14){$-\ell$}
    \put(8,115){\line(1,0){4}}
    \put(0,112){$\ell$}
    \put(70,0){(a)}
    \put(190,0){(b)}
    \put(310,0){(c)}
  \end{picture}
  \caption{$h < h_{\ell}$ for (a). $h > h_{\ell}$ for (b) and (c).} 
  \label{fig:dis}
  \end{center}
  \vspace{1cm}
\end{figure}

Now let us consider the case where the radius $h\lambda N$ 
of the spherical D2-brane or fuzzy two-sphere grows bigger than 
the half of the circumference $\ell$ of the $x^3$-circle.
Now $\ell$ is fixed as
\begin{alignat}{3}
  \ell = h_{\ell} \lambda N .
\end{alignat}
When $h$ is smaller than $h_\ell$, the spherical configuration is the
stablest with the potential energy $NT_0 - \frac{2}{3}NT_0 
(h^2\lambda N)^2$ (Fig.\ref{fig:dis}(a)). As $h$ is approaching to $h_\ell$, 
the tachyonic modes will appear from the open strings stretched between 
the adjacent surfaces. And when $h$ grows bigger than $h_\ell$, 
the tachyon condensation will occur. Though we cannot trace this process,
we know that D2-brane must emit some D0-branes to make its size small. 
Suppose that $M$ D0-branes are emitted during the tachyon condensation. 
Then $\ell$ and $h\lambda (N\m M) \,( < \ell)$ are obtained, 
so we can calculate $X$ or $R_N$ in principle and predict the shape 
of the torus-like dielectric D2-brane after the tachyon 
condensation(Fig.\ref{fig:dis}(b)).
Of course the torus-like dielectric D2-brane is energetically unstable 
since the potential energy of it 
is bigger than that of the spherical dielectric one with the radius
$h\lambda (N-M)$ (Fig.\ref{fig:dis}(c)). 
The difference between these two potential energies
is, however, less than $NT_0 (h^2\lambda (N\m M))^2$.
Thus if $(h^2\lambda (N\m M))^2$
is small enough, the torus-like dielectric D2-brane can be regarded as
a meta-stable state.

The K-theory analysis leads that a non-BPS D-string is a spharelon
in string theory\cite{HHK}.
In our analyses, the meta-stability of the torus-like D2-brane
came from the existence of the constant R-R flux, which does not 
appear in the K-theory analysis. However,
the fact that the torus-like D2-brane is meta-stable suggests that 
these solutions are puffed analogue of the non-BPS D-string.
On the other hand, the non-BPS D-string may emit only one D0-brane.
It is interesting if we can shrink the cycle of the torus-like D2-brane
and compare the properties of these two.
It is also a challenging task to ask that a stable non-BPS D0-brane
in type I superstring theory can be constructed from circling
a D-string with non-trivial ${\mathbb Z}_2$ Wilson line.

\vspace{1cm}

\vspace{1cm}
\section{Acknowledgments}

The author would like to thank Hiroshi Kunitomo for useful discussions
and careful reading of the manuscript, Hidetoshi Awata, Kenji Hotta, 
Tadakatsu Sakai, Masafumi Fukuma and Masao Ninomiya for useful 
conversations or encouragements.

\appendix

\vspace{1cm}
\section{The Properties of the Elliptic Integrals}
\label{sec:elli}

In this appendix, the properties of the incomplete elliptic
integrals of the first kind $F(\psi,k)$ and second kind $E(\psi,k)$ 
are shortly explained. The definition of $F(\psi,k)$ and $E(\psi,k)$
are given by
\begin{alignat}{3}
  F(\psi,k) \equiv \int_0^\psi \!\! d\phi \frac{1}{\sqrt{1-k^2 \sin^2 \phi}}
  ,\quad
  E(\psi,k) \equiv \int_0^\psi \!\! d\phi \sqrt{1-k^2 \sin^2 \phi} .
\end{alignat}
$K(k) \equiv F(\frac{\pi}{2},k)$ and $E(k) \equiv E(\frac{\pi}{2},k)$ are
called the complete elliptic integrals of the first and second kinds
respectively. And at $k \!=\! 0,1$,
\begin{alignat}{3}
  K(0) = \frac{\pi}{2} \;,\quad K(1) = \infty \;,\quad
  E(0) = \frac{\pi}{2} \;,\quad E(1) = 1 .
\end{alignat}
A formula
\begin{alignat}{3}
  \int_0^\frac{\pi}{2} \!\! d\phi \Big( \sqrt{1-k^2 \sin^2 \phi} \Big)^3
  = \frac{2(2-k^2)}{3} E(k) - \frac{1-k^2}{3} K(k),
\end{alignat}
is used in eq.(\ref{eq:ene}).

\end{document}